# Beam Dynamics and Layout

*A.M. Lombardi*
CERN, Geneva, Switzerland

**Abstract**
In this paper, we give some guidelines for the design of linear accelerators, with special emphasis on their use in a hadron therapy facility. We concentrate on two accelerator layouts, based on linacs. The conventional one based on a linac injecting into a synchrotron and a all-linac solution based on high gradient high frequency RF cavities.

## 1 Introduction

Accelerators for use in medical applications should produce a proton beam with an energy of about 250 MeV, a carbon-ion beam with an energy of about 450 MeV/u, or both, with an average current of around 30 nA and a peak microbunch current of 500 µA. Ideally, a medical accelerator should be compact, cheap, reliable, easy to operate, and modular.

There are two established designs of medical accelerators that need linacs, one based on a linac injecting into a synchrotron and another (all-linac solution) based on linear accelerators. Their respective merits and drawbacks are not for discussion here, the main difference being the possibility of changing the beam energy from pulse to pulse at a very rapid rate if the all-linac solution is chosen, thus considerably reducing the treatment time. In this paper we will discuss the choices for the layout of both options, which are sketched in Figs. 1 and 2.

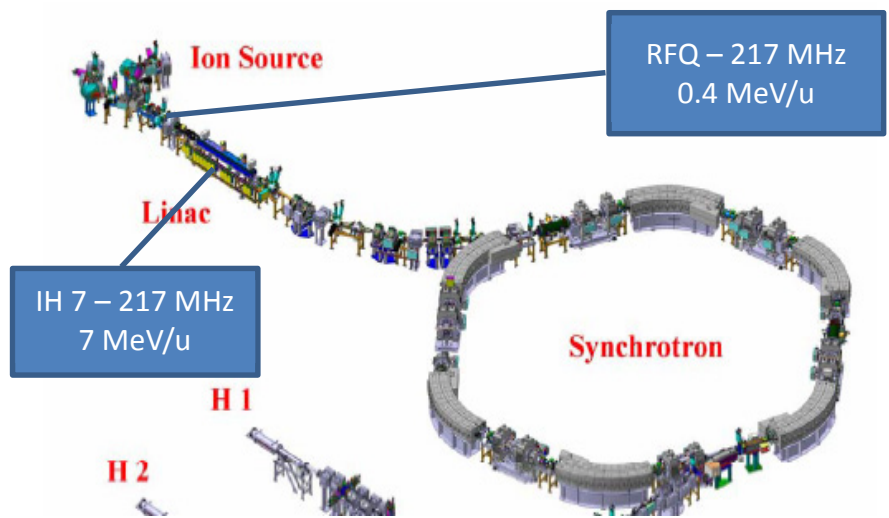

**Fig. 1:** Sketch of a medical facility based on a linac (5 m long) injecting into a synchrotron (10 m radius): typically the linac is composed of a Radio Frequency Quadrupole (RFQ) and a Interdigital H (IH) structure operating at a frequency of around 200 MHz with injection into the synchrotron at an energy of 7 MeV/u. This scheme can be used with protons and carbon ions, coming from two different sources funnelled into the RFQ at the low-energy end.

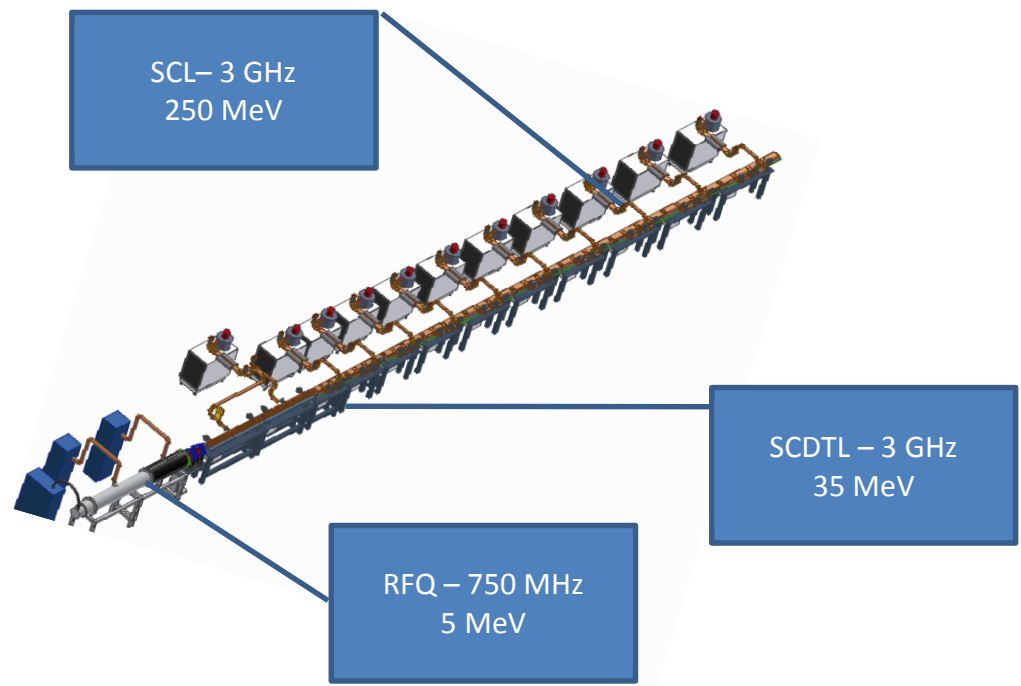

**Fig. 2:** Sketch of a medical facility based on high-frequency linacs, about 30 m in length, composed of a 750 MHz RFQ to bring a proton beam to 5 MeV and injecting into a Side-Coupled Drift Tube Linac (SCDTL) at 3 GHz. The final energy is reached with a standard Side-Coupled Linac (SCL) at 3 GHz. Such a facility can provide extremely fast (200 Hz) pulse-to-pulse energy and current variability for a proton beam.

## 2 Fundamentals of linear accelerators for medical applications

In this paper, we discuss guidelines for the design of a linear accelerator for medical applications, bearing in mind the two schemes shown above. For simplicity, and justified by the fact that the current in medical accelerators is rather modest, we treat the longitudinal and transverse planes separately.

### 2.1 Longitudinal plane: bunching and acceleration

A proton or carbon beam generated by a particle source is continuous on the scale of the radio frequency (RF) used in a linac. As it is not possible to transfer energy to a continuous beam by means of an RF field, it is necessary to prepare a beam from the source for RF acceleration. The section that achieves this is called a pre-injector. It is generally composed of an RF quadrupole, which also has the function of accelerating the beam to the input energy of the injector and of shaping the longitudinal emittance to match it to the acceptance of the injector. This manipulation of the beam to prepare for RF acceleration should be done whilst controlling losses and minimizing emittance growth. The pre-injector typically increases the energy of the beam to a few MeV over a few metres and is not very efficient.

The main function of the pre-injector is to bunch the beam on the scale of the wavelength. This operation is done by generating a velocity spread in the beam by passing it through an RF cavity and then letting the beam distribute itself around the particle with the average velocity. We can distinguish two extreme types of bunching: discrete and adiabatic.

Discrete bunching is shown in the sketch in Fig. 3, where the longitudinal phase space (phase–energy space) is shown at five different steps, starting from the top left. A beam from an ion source is continuous (1); this beam passes through an RF cavity, with the result that some particles are accelerated and some are decelerated (2). Notice that the average energy is not changed. After passing through the

RF cavity, the beam has a velocity distribution (2), which induces a change in the relative positions of the particles (3, 4). At the moment 5 the slowest, the average, and the fastest particle are at the same physical location: they are grouped around the average particle, as can be appreciated from the phase histogram of part 5 of the figure, shown in part 6.

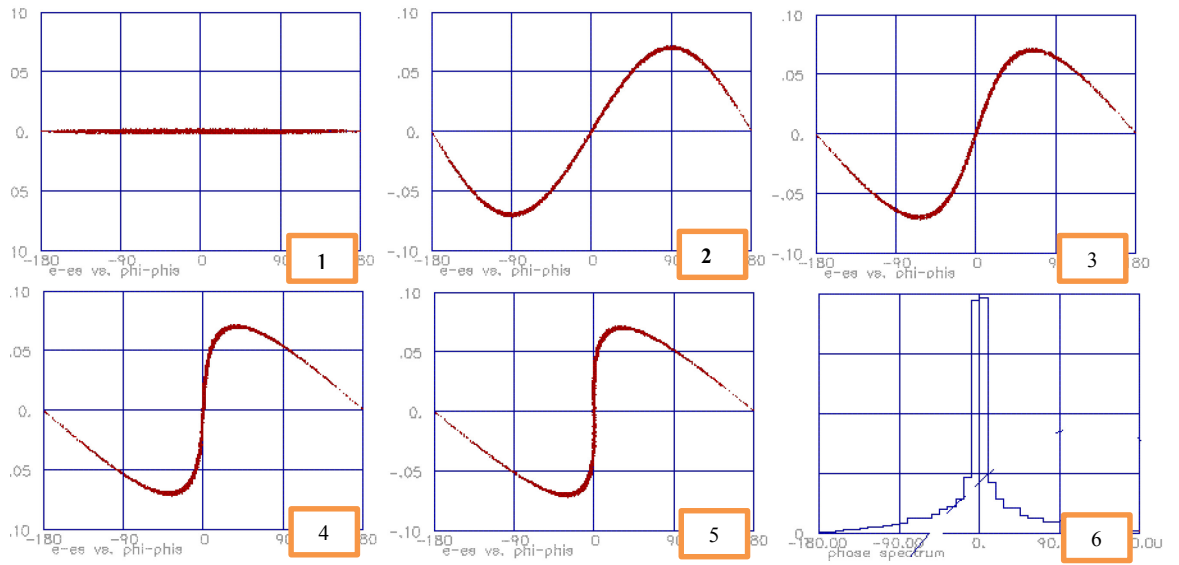

**Fig. 3:** Sketch of a discrete bunching process

The efficiency of discrete bunching is about 50%; namely, about 50% of the initial beam can be accelerated further by a chain of RF cavities. With a particularly well-designed system which includes higher harmonics of the frequency, an efficiency of 60–70% can be achieved. Nevertheless, a substantial fraction of the beam is lost. To improve the efficiency a different approach to bunching, called adiabatic bunching, can be used. Adiabatic bunching entails continuous bunching at a very low voltage, which gives the beam time to wrap around the synchronous phase. The concept is to generate a velocity spread continuously with a small longitudinal field and perform the bunching over several oscillation in the phase space (up to 100!). This allows better capture around the stable phase, achieving up to 95% capture. Adiabatic bunching is performed in the first few sections of an RFQ by slowly increasing the depth of the modulation along the structure, thus making it possible to bunch the beam smoothly and prepare it for acceleration. The beam needs to be kept bunched during the whole acceleration phase, i.e., it is necessary to provide a longitudinal restoring force. This is done by accurately choosing the phase of acceleration, as shown in Fig. 4: particles with less energy than the average should see a slightly higher accelerating field than the particle with the average energy, and the opposite should be true for particles with more energy than the average. This concept is known as the principle of phase stability.

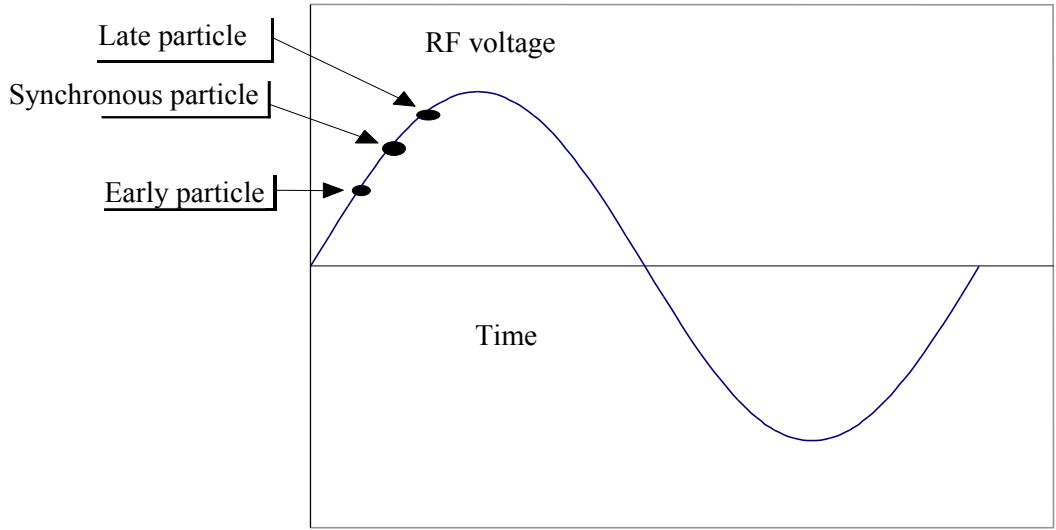

**Fig. 4:** Sketch of the principle of phase stability

Once the beam has been bunched and is ready to be accelerated, the average velocity of the beam will change as it traverses RF cavities. In the design phase, the layout is calculated around the average particle of the beam, often called the synchronous particle. The synchronous particle is the particle (possibly fictitious) used to calculate and determine the phase along the accelerator. It is the particle whose velocity is used to determine the synchronicity with the electric field. The idea is to design for the synchronous particle and provide longitudinal focusing so that other particles will perform small oscillations around it and remain bound to the centre of the bunch.

The length of each accelerating element determines the time at which the synchronous particle enters or exits an RF cavity, so for a given cavity length there is an optimum velocity (or beta = velocity/speed of light) such that a particle travelling at that velocity passes through the cavity in half an RF period. The difference in time of arrival between the synchronous particle and a particle travelling at a speed corresponding to the geometrical beta (the velocity of a particle which would traverse the cavity in half the RF period) determines the phase difference between two adjacent cavities. We can therefore adjust the phase between two adjacent RF cavities by changing the length of one of the cavities. In a synchronous structure, the geometrical beta is always equal to the synchronous-particle beta and each cell is different. A synchronous structure provides the best possible longitudinal beam dynamics and allows full control of the longitudinal phase space, but it implies that each cavity is different. For medical applications, it is possible to lift this constraint after the beam has become energetic enough (beta = 20%), to allow some standardization of the cavity length. To simplify construction and contain costs, cavities are not individually tailored to the evolution of the beam velocity but are constructed in blocks of identical cavities (tanks). Several tanks are fed by the same RF source. This simplification implies a 'phase slippage', i.e., motion of the centre of the beam around a stable phase. The phase slippage is proportional to the number of cavities in a tank, and it must be carefully controlled for successful acceleration.

Let us now turn to acceleration. We shall describe the motion of a particle in the longitudinal phase space and establish a relation between the energy and phase of the particle during acceleration.

If we write the energy gain of the synchronous particle as
$$\Delta W_s = qE_0 LT\cos(\varphi_s), \tag{1}$$

where the symbols are defined in the lecture "Overview of linacs" in these proceedings, then the energy gain of a particle with phase φ is

$$\Delta W = qE_0 LT \cos(\varphi),  \qquad (2)$$

and assuming a small phase difference $\Delta\varphi = \varphi - \varphi_s$, we can write the following equations for the energy and phase:

$$\frac{d}{ds}\Delta W = qE_0 T \cdot [\cos(\varphi_s + \Delta\phi) - \cos\varphi_s],  \qquad (3)$$

$$\frac{d}{ds}\Delta\varphi = \omega\left(\frac{dt}{ds} - \frac{dt_s}{ds}\right) = \frac{\omega}{c}\left(\frac{1}{\beta} - \frac{1}{\beta_s}\right) \cong -\frac{\omega}{\beta_s c}\frac{\Delta\beta}{\beta_s} = -\frac{\omega}{mc^3 \beta_s^3 \gamma_s^3}\Delta W .  \qquad (4)$$

The expressions above are equations for the canonically conjugate variables phase and energy, with Hamiltonian (total energy of oscillation)

$$\frac{\omega}{mc^3\beta_s^3\gamma_s^3}\left\{\frac{\omega}{2mc^3\beta_s^3\gamma_s^3}(\Delta W)^2 + qE_0 T[\sin(\varphi_s + \Delta\varphi) - \Delta\varphi\cos\varphi_s - \sin\varphi_s]\right\} = H .  \qquad (5)$$

For each $H$, we have different trajectories in the longitudinal phase space. The equation of the separatrix (the line that separates stable from unstable motion) is

$$\frac{\omega}{2mc^3\beta_s^3\gamma_s^3}(\Delta W)^2 + qE_0 T[\sin(\varphi_s + \Delta\varphi) + \sin\varphi_s - (2\varphi_s + \Delta\varphi)\cos\varphi_s] = 0 ,  \qquad (6)$$

from which we can deduce that the maximum energy excursion of a particle moving along the separatrix is

$$\Delta\hat{W}_{max} = \pm 2\left[\frac{qmc^3\beta_s^3\gamma_s^3 E_0 T(\varphi_s\cos\varphi_s - \sin\varphi_s)}{\omega}\right]^{\frac{1}{2}} .  \qquad (7)$$

This is a very important and useful expression that gives the energy acceptance of an accelerator depending on the field level and the accelerating phase. The longitudinal acceptance of an accelerator has a characteristic shape, similar to a golf club. Particles falling into this area in phase space can be successfully accelerated. A practical example is shown in Fig. 5.

When we accelerate on the rising part of the positive RF wave, we have a longitudinal force which keeps the beam bunched. The force (of harmonic-oscillator type) is characterized by the longitudinal phase advance, expressed as

$$k_{0l}^2 = \frac{2\pi qE_0 T \sin(-\varphi_s)}{mc^2 \beta_s^3 \gamma^3 \lambda}\left[\frac{1}{m^2}\right],  \qquad (8)$$

with the equation for the phase resulting in

$$\frac{d^2\Delta\varphi}{ds^2} + k_{0l}^2\left(\Delta\varphi - \frac{\Delta\varphi^2}{2\tan(-\varphi_s)}\right) = 0 .  \qquad (9)$$

An exception is the KONUS beam dynamics used in the IH structure [1], where the particle beam is purposely accelerated outside the area of stability, i.e., outside the separatrix. This dynamics is very convenient for accelerating efficiently, but it requires periodic rebunching sections.

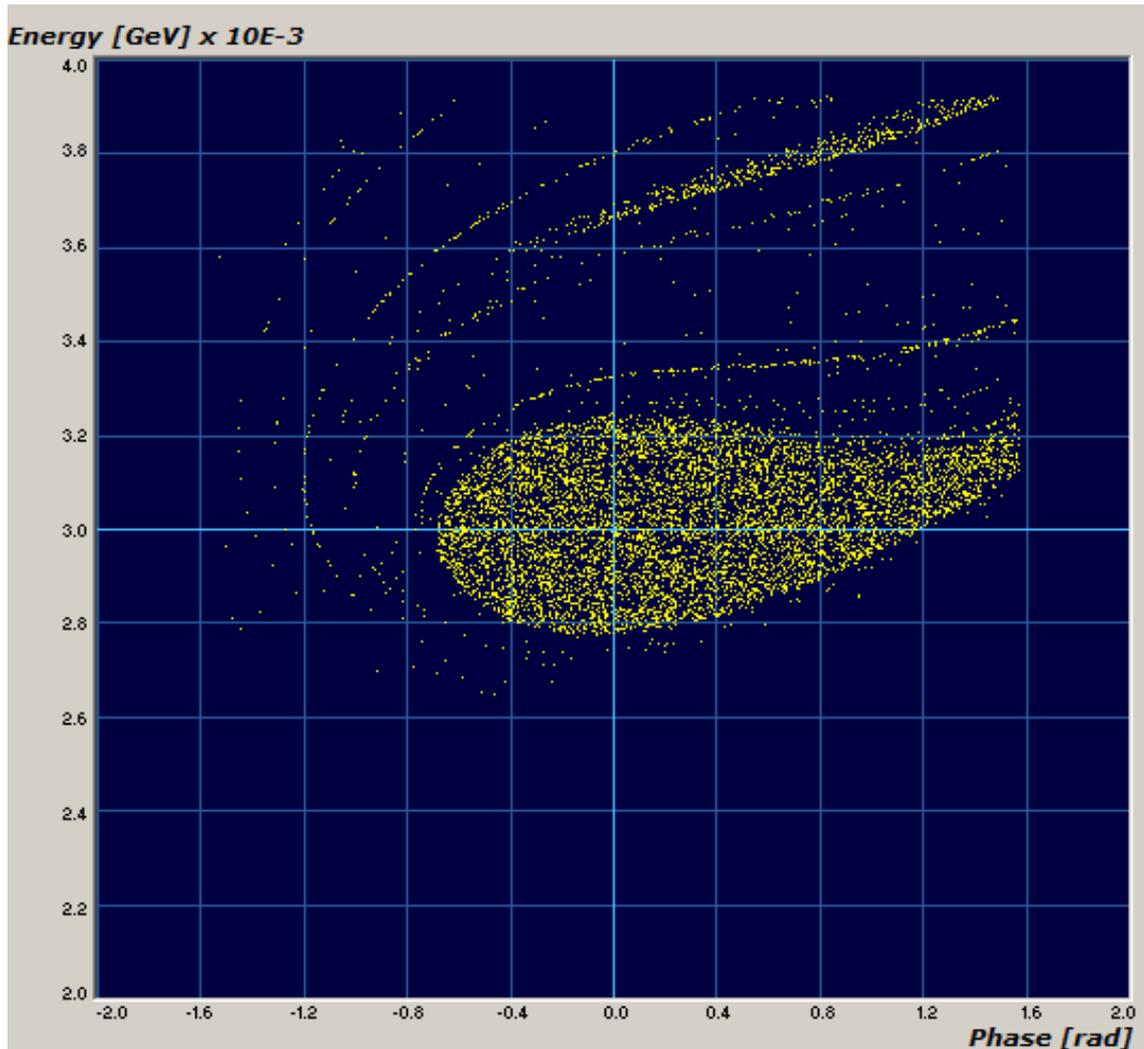

**Fig. 5:** Longitudinal acceptance of the CERN LINAC4 DTL (352 MHz, 3–50 MeV)

## 2.2 Transverse plane: focusing

All along the accelerator we need to provide a transverse force, to keep the beam confined transversely. Ideally we should apply a force towards the beam axis proportional to the distance from the axis, a linear force which would keep the beam confined without degrading the emittance. In the case of medical accelerators, the microbunch current is very low and space charge effects are generally negligible, thus simplifying the layout of the optics and allowing weak focusing. The only choice that can be made with respect to the focusing strength depends on the bore radius of the cavities (stronger focusing, smaller radius) and on the accelerating phase. As we will see in the following, at low energy there is substantial coupling between the transverse and longitudinal planes caused by the transverse defocusing RF effects, which is implied by the choice of a stable restoring force in the longitudinal plane to keep the beam bunched.

There are two types of focusing force: electric focusing and magnetic focusing. Electric focusing is independent of the particle velocity and is therefore generally used at the low-energy end. We have talked extensively about the electric focusing force in the lecture "Overview of linacs" in these proceedings.

A magnetic quadrupole (electromagnetic or composed of a special arrangement of permanent magnet material) provides a field $B$ that can be expressed as

$$\begin{cases} B_x = G \cdot y \\ B_y = G \cdot x \end{cases}, \qquad (10)$$

where $G$ is the quadrupole gradient, i.e., the field on the pole tip divided by the quadrupole bore radius. A charged particle (with charge $q$) travelling at a velocity $v$ through this field experiences a force proportional to its distance $(x, y)$ from the axis, which can be written as

$$\begin{cases} F_x = -q \cdot v \cdot G \cdot x \\ F_y = q \cdot v \cdot G \cdot y \end{cases}. \qquad (11)$$

The force in the expression above is focusing in the $x$ plane and defocusing in the $y$ plane for a positive gradient $G$.

In order to keep the beam confined along the accelerator axis, it is therefore necessary to interlace a series of quadrupoles of alternate polarity. This arrangement of quadrupoles is called a FODO channel. The beam dynamics in a FODO channel is shown in Fig. 6. The beam envelope in one plane is shown at characteristic locations along the channel, and the corresponding orientation of the beam emittance is shown below it. It should be appreciated that after one full focusing period (positions 1 and 7) the beam phase space is identical to what it was before. Such a channel can be extended throughout the whole accelerator, taking into account the fact that its length needs to be progressively adapted to the changing velocity of the beam.

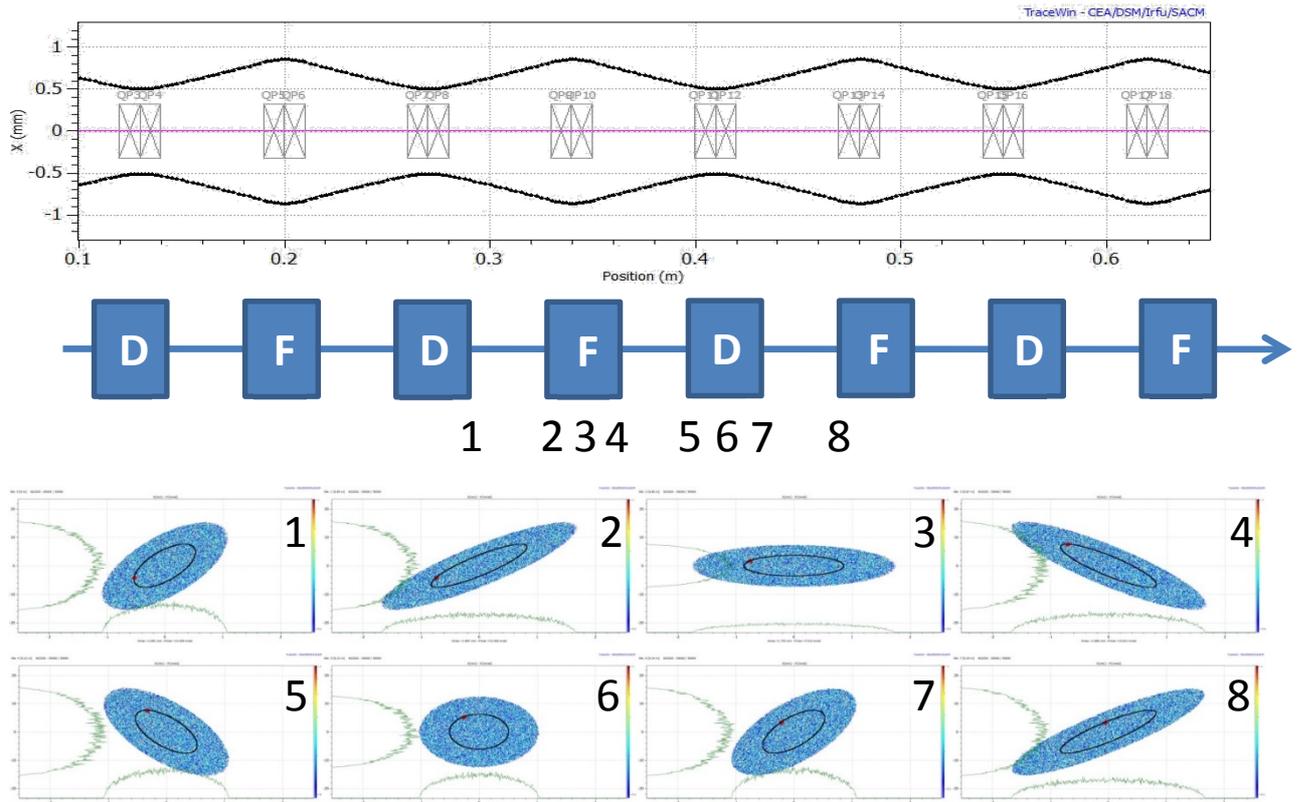

**Fig. 6:** Beam envelope (top) and beam phase space (bottom) in a FODO channel

We can write the solution of the equation of motion in a periodic channel as

$$X(s) = \sqrt{\varepsilon \beta(s)} \cos(\sigma_{0t}(s)), \qquad (12)$$

where $\varepsilon$ is the beam emittance, $\beta$ is a periodic function with the periodicity of the focusing period, and $\sigma$, the transverse phase advance, is a measure of the strength of the focusing channel.

NB: in Eqs. (12) and (14), $\beta$ is *not* the relativistic $\beta$.

In medical linacs, the overall force balance in the transverse plane is given by the static quadrupole focusing and the RF defocusing and can be expressed via the transverse phase advance at zero current,

$$\sigma_{0t} = \sqrt{\frac{\theta_0^4}{8\pi^2} + \Delta_{rf}} \ . \tag{13}$$

The first term on the right-hand side depends on the strength of the quadrupoles and the particle velocity according to

$$\theta_0^2 = \frac{qG\lambda^2 N^2 \beta \chi}{m_0 c \gamma}, \tag{14}$$

where $G$ is the magnetic quadrupole gradient (in units of T/m), $N$ is the number of magnets in a period, and $\chi$ is as follows:

for $+-$ ($N = 2$):

$$\chi = \frac{4}{\pi}\sin\left(\frac{\pi}{2}\Gamma\right); \tag{15}$$

for $++--$ ($N = 4$):

$$\chi = \frac{8}{\sqrt{2}\pi}\sin\left(\frac{\pi}{4}\Gamma\right), \tag{16}$$

where $\Gamma$ is the quadrupole filling factor (the quadrupole length relative to the period length).

The expression above allows one to determine the quadrupole gradient necessary to limit the beam size to a given value for a given quadrupole configuration and a given beam energy.

The RF defocusing, represented by the term $\Delta_{rf}$, comes from the varying electric field in the RF cavities and is a consequence of the choice of the accelerating phase according to the principle of phase stability. If we write the Maxwell equation

$$\nabla \cdot E = 0 \tag{17}$$

as

$$\frac{\partial E_x}{\partial x} + \frac{\partial E_y}{\partial y} + \frac{\partial E_z}{\partial z} = 0, \tag{18}$$

we can see that a longitudinal restoring force

$$\frac{\partial E_z}{\partial z} > 0 \tag{19}$$

implies a transverse defocusing

$$\frac{\partial E_x}{\partial x} + \frac{\partial E_y}{\partial y} < 0 \tag{20}$$

with an intensity equal to half of the longitudinal phase advance:

$$\Delta_{\rm rf} = \frac{1}{2}\sigma_{0l}^2 = \frac{1}{2}\beta^2\lambda^2 k_{0l}^2 = \frac{\pi q \lambda N^2 E_0 T \sin\phi_s}{m_0 c^2 \beta \gamma^3} \ . \tag{21}$$

The RF defocusing is a phase-dependent defocusing term which is more important at the lowest energies. It is often the parameter that determines the focusing layout at energies up to 5 MeV and frequencies of the order of 200 MHz.

## 3  Putting it all together: a rough guide

In this final section we attempt to give some guidelines for the design of a linear accelerator for medical applications. Let us assume we have available a collection of RF cavities to increase the beam energy, such as an RFQ, an IH structure, a drift tube linac, a side-coupled linac, and some hybrid structure. Assume we have available solenoids, electromagnetic quadrupoles, and permanent magnet quadrupoles to keep the beam volume confined. Assume also that we have a green-field site on which to design an accelerator. The accelerator designer has to make some (difficult) starting choices which determine the layout of the accelerator. These choices are not generally straightforward, as each choice has different implications and there is not generally only one "right choice". In the following we list a series of questions that are intended to trigger discussions and thoughts towards making an informed choice.

### 3.1  First basic questions and choices

*Which frequency?* The first and foremost important choice is the operating frequency. There are some standardized frequencies for an accelerator for which power sources are available. For an accelerator that is to be built, it is advised to choose a frequency for which a power source exists. The range of available frequencies is still rather large and the choice of the frequency has strong implications for the transverse size of the accelerator, the transverse acceptance, and the maximum accelerating field and duty cycle. The higher the frequency, the more compact the accelerator, but at the same time the smaller the acceptance. It can also be envisaged, for an all-linac solution, that one may have a frequency jump during the acceleration, thus optimizing the frequency for the energy range that is to be used.

*At what energy does one make the transition between a TE structure (RFQ) to a TM structure (e.g., DTL)?* This point should be evaluated when the design is already somewhat advanced, ideally after two structures overlapping in energy have been designed and several combinations of transition energies have been tried. In general, transitions are points of weakness in accelerators, especially at the low-energy end, and should be designed carefully (with enough variable elements to accommodate errors). A transition below the threshold for activation of neutron production in copper (about 3–5 MeV) is a good choice.

As we have seen, medical accelerators should be compact, so ideally we would like to chain as many RF cavities as possible together and reduce to a minimum the number of focusing elements. But *what is a sensible minimum number of focusing elements?* We should look at the beam size in a FODO channel: the maximum beam size increases as the length of the FODO period increases. So the smaller the number of quadrupoles, the larger the beam size and the larger the bore radius that we need in the RF cavities. A larger bore radius implies more power for the same accelerating gradient, so the answer to this question is a balancing exercise between real estate gradient, the bore aperture (and hence RF efficiency), and the transverse acceptance (source performance).

As we have seen, it is convenient to standardize the structures and have RF cavities that are all of the same length. It is economical to have as many RF gaps per cavity as possible, but *what is the maximum number of gaps that we can put together?* In this case also, we are faced with a balancing exercise between the acceptable phase slippage and RF power optimization and distribution, with implications for the longitudinal acceptance, i.e., the quality of the pre-injector.

*What transverse acceptance should the accelerator provide?* This choice is dictated mainly by cost. A large acceptance makes the operation of the accelerator easier and implies less stringent alignment and machining tolerances. On the other hand, a large acceptance means a larger bore radius and therefore less efficient acceleration.

*Variable or fixed focusing system?* The choice is between a fixed focusing system (using permanent magnet quadrupoles) and a more costly variable system. In general, for an all-linac solution, a fixed focusing system should be chosen because it is economical, it simplifies operation, and reduces the size of the accelerator. A variable focusing system is strictly needed only in transitions between structures and if different ion species need to be accommodated in the same accelerator.